\documentstyle[12pt,epsf]{article}
\def\beq{\begin{equation}}
\def\eeq{\end{equation}}
\def\bea{\begin{eqnarray}}
\def\eea{\end{eqnarray}}
\def\bef{\begin{figure}}
\def\enf{\end{figure}}
\def\S{{\bf S}}
\def\A{{\bf A}}

\def\C{{\bf C}}
\def\CP{{\bf CP}}

\def\Z{{\bf Z}}
\def\R{{\bf R}}

\def\T{{\bf T}}

\def\w{{\bf w}}
\def\r{{\bf r}}
\def\y{{\bf y}}
\def\v{{\bf v}}
\def\u{{\bf u}}

\def\CN{{\cal N}}

\def\ba{\begin{array}}
\def\ea{\end{array}}
\def\bce{\begin{center}}
\def\ece{\end{center}}

\begin {document}
\begin{titlepage}
\hfill
\vbox{
    \halign{#\hfil         \cr
           hep-th/0205067 \cr
           HU-EP-02/20 \cr
           } 
      }  
\vspace*{20mm}
\begin{center}
           {\Large {\bf
Orbifolds, Penrose Limits and Supersymmetry Enhancement} \\ }
\vspace*{15mm}
\vspace*{1mm}
{Kyungho Oh$^a$ and Radu Tatar$^b$}

\vspace*{1cm}

{\it $^a\ $Department of Mathematics, Computer Science, Physics and Astronomy,
\\University of Missouri at St Louis, St Louis, USA}\\
{\tt {e-mail: oh@arch.umsl.edu}}\\
\vspace*{2mm}
{\it $^b\ $Institut fur Physik, Humboldt Universitat\\
Berlin, 10115, Germany}\\
{\tt {e-mail: tatar@physik.hu-berlin.de}}\\

\vspace*{1cm}
\end{center}

\begin{abstract}

We consider supersymmetric PP-wave limits for different $\CN=1$ orbifold geometries of the five sphere $\S^5$ and
the five dimensional Einstein manifold $\T^{1,1}$. As there are several interesting ways to take the Penrose
limits, the  PP-wave geometry can be either maximal supersymmetric ${\cal N}=4$ or half-maximal supersymmetric
${\cal N}=2$. We discuss in detail the cases $AdS_5\times \S^5/\Z_3$, $AdS_5\times \S^5/(\Z_m \times \Z_n)$ and
$AdS_5\times \T^{1,1}/(\Z_m \times \Z_n)$ and we identify the gauge invariant operators which correspond to
stringy excitations for the different limits.

\end{abstract}

\vskip 1in

\leftline{May 2002}

\end{titlepage}

\newpage
\section{Introduction}
The duality between open strings and closed strings has been explored extensively over the last years. One
important example is the AdS/CFT conjecture between the ${\cal{N}}=4$ field theory and type IIB strings on
$\bf{AdS_5 \times S^5}$ \cite{mal,gkp1,wi1}. The conjecture has been generalized to orbifolds of  $\bf{S^5}$
\cite{ks,law} and to conifolds \cite{kw}. The supergravity limit of the string has been mainly considered so far
because of the difficulties in quantizing strings in the presence of RR-fluxes. On the other hand, another
maximal supersymmetric background, the pp-wave, has been discussed recently in \cite{blau} and string theory on
the pp-wave is an  exactly solvable model, where one can identify all the string oscillators \cite{metsaev}.

The pp-wave solution appear as a Penrose limit of the   $\bf{AdS_5 \times S^5}$ solution \cite{blau,bmn} so it
can be used to obtain information about the AdS/CFT correspondence. The authors of \cite{bmn} have extended the
AdS/CFT conjecture to the case of strings moving on a pp-wave backgrounds where the corresponding field theory
operators are the ones with high R charge and in this case the field theory  describe not only the supergravity
but also the full closed string theory.

The idea of \cite{bmn} has been extended in many directions \cite{papers, z2, ikm, go, pando, mukhi, ali, bmn1}.
The direction we are pursuing in this work was initiated in a series of papers \cite{z2, ikm, go, pando, mukhi,
ali, bmn1} and involves geometries more complicated than the $\bf{S^5}$. Especially interesting are the cases of
orbifolds of $\bf{S^5}$ or conifolds where one can take two kinds of pp-wave limits, one which preserves the
supersymmetry and the other one which enlarges the supersymmetry. As discussed in \cite{blau}, if one takes the
Penrose limit on directions orthogonal to the orbifolding direction, then we expect to get  the same amount of
supersymmetry, but a Penrose limit along the orbifolding direction will get an increase of supersymmetry. One
example of the second type was described in  \cite{ikm,go,pando} for the case of D3 branes at a conifold
singularity, where the Penrose limit gives a maximal supersymmetric solution. In this case we expect a
supersymmetry enhancement in field theory, from ${\cal{N}}=1$ to ${\cal{N}}=4$ and the relevant ${\cal{N}}=1$
multiplets which give rise to an ${\cal{N}}=4$ multiplet have been identified. In \cite{mukhi,ali} a similar
discussion has been developed for the supersymmetry enhancement from ${\cal{N}}=2$ to ${\cal{N}}=4$ in the case
of $\S^5/\Z_k$.

In the present work we study  the supersymmetry enhancement in the Penrose limit for several examples of
orbifolds. As  only the infinitesimal neighborhood of  the null geodesic is probed in the pp-wave limit, the
orbifold action disappears unless it is considered locally around the null geodesic. In other words, the
orbifolding action also changes in the pp-wave limit.  Thus, in general, it is not possible to build duals to
string oscillators in the Penrose limit from gauge invariant operators of the original orbifold theory. We have
found that, in the Penrose limit, one needs to consider operators from the covering space of the original space.
We also comment on anomalous dimensions and correlation function for the
orbifold theories and on  the interpretation as a limit of a DLCQ theory
with the light-cone momentum $p^+$ fixed.

In section 2 we will describe examples of ${\cal N} = 1$ orbifolds of $\bf{S^5}$. The first model is
$\bf{S^5/Z_3}$ whose Penrose limit was outlined in \cite{go} for which we describe the string/field theory
matching. As a second example we consider different boostings for the $\S^5/(\Z_k \times \Z_l)$ orbifold which
can give an enlargement of supersymmetry from ${\cal N} = 1$ to ${\cal N} = 2$ or ${\cal N} = 4$. In section 3 we
consider Penrose limits of $\T^{1,1}/(\Z_k \times \Z_l)$ along the fixed circles of the quotienting action.

\section{${\cal{N}} = 1$ orbifolds of $\S^5$}
\subsection{Review of the $AdS_5 \times \S^5$ result}
We start with a brief review of the result of \cite{bmn}, pointing out
the features which we expect to get from the orbifold discussion.

Consider $AdS_5 \times \S^5$ where the anti-de-Sitter space $AdS_5$ is represented as a universal covering of a
hyperboloid of radius $R$ in the flat $\R^{2,4}$ and a sphere $\S^5$ of radius $R$ in the flat space $\R^{0,6}$.
One may regard the $AdS_5$ (resp. $\S^5$) as a foliation of a time-like direction and a three sphere $\Omega_3$.
(resp. a circle parameterized by $\psi$ and a three sphere $\Omega_3'$.) Then the induced metric on $AdS_5\times
\S^5$ becomes \bea ds^2= R^2\left[ -dt^2 \cosh^2 \rho + d\rho^2 + \sinh^2\rho d\Omega_3^2 + d \psi^2 \cos^2
\theta + d \theta^2 + \sin^2 \theta d \Omega_3^{'2}\right]. \eea

One now considers the pp-limit by boosting  along the $\psi$ direction around $\rho =0$. The metric in this limit
can be obtained by taking $R \to \infty$ after introducing coordinates \bea \label{lcc} x^{+} = \frac{1}{2} (t +
\psi), x^{-} = \frac{R^2}{2} (t - \psi) \eea  and rescaling $\rho  = r/R, \theta = y/R$ as follows:
\bea\label{ppwaven4} ds^2 = -4 dx^{+} dx^{-} - (\r\cdot \r + \y \cdot \y)  dx^{+ 2} + d \y^2 + d\r^2\eea where
$\y$ and $\r$ parameterize points on $\R^4$. Only the components of the RR
5-form $F$ with a
plus index survive in this limit.

The energy is given by $E = i \partial_{t}$
and the angular momentum in
the direction $\psi$ is $J = - i \partial_{\psi}$ and the latter is seen as
a generator that rotates a 2-plane inside the original $\bf{R^6}$.

In terms of the dual ${\cal{N}} = 4$ theory, the energy $E$ is related to
the conformal weight $\Delta$ and the angular momentum to the R-charge.
As discussed in \cite{bmn}, the relation between the oscillations of the
string in the pp-wave geometry (\ref{ppwaven4}) and the field theory
quantities is
\bea
(\Delta - J)_n = \sqrt{1 + \frac{4 \pi g N n^2}{J^2}}
\eea
where N stands for the rank of the gauge theory and g is the string
coupling constant. The vacuum has $\Delta - J = 0$.

In the ${\cal{N}} = 4$ field theory, the interpretation of the string vacuum and of the string oscillators is
made in terms of the gauge invariant operators. Consider the $\CN= 4$ multiplet in terms of a triplet of
 $\CN = 1$ multiplets, denoted by $Z, Y^1, Y^2$, the dimension of
each field being 1. The complex field $Z$ is on the directions whose rotation generator is $J$, so the value of
$J$ for the field $Z$ is 1, therefore for the field $Z$ we have $\Delta - J = 0$. The other fields, $Y^1, Y^2$
(and their complex conjugates $\bar{Y}^1, \bar{Y}^2$) have $J=0$
and $\Delta - J = 1$.

We can
proceed to compare the stringy results with the field theory results,
the string vacuum is given by $\mbox{Tr}[Z^J]$ and the stringy oscillators
are given by inserting $Y^1, Y^2, \bar{Y}^1, \bar{Y}^2$ , i.e. the
operators:
\bea
\mbox{Tr}[Z^{J-1}] Y^i, ~ \mbox{Tr}[Z^{J-1}] \bar{Y}^i, i=1,2
\eea

We can also have gauge invariant operators $\mbox{Tr}[Z^{J-1}] \bar{Z}$,
$\mbox{Tr}[Z^{J-2}] Y^i \bar{Y^j}$ , etc, but in \cite{bmn} arguments have
been given that such operators will get infinite mass.

\subsection{String Oscillators in the pp-limit of the $AdS_5 \times \S^5/\Z_3$}
The geometry $AdS_5 \times \S^5/ \Z_3$ is obtained as a near horizon geometry of $N$ D3 branes placed at a
$\C^3/\Z_3$ orbifold.  The generator $g$ of $\Z_3$ acts on $\C^3$ by \bea g \cdot (z_1, z_2, z_3) \to (\omega
z_1, \omega z_2, \omega z_3), ~~~\omega^3 =1. \eea

We consider the boosting along the direction of the orbifolding which has been studied in \cite{go}. We need to
consider a metric for the 3-fold covering of $\S^5$. As \cite{go}, it is convenient to consider $\S^5$ as a Hopf
fibration over $\CP^2$. The metric could be written as: \bea \label{metricz3} ds^2 = (3d\psi + A)^2 +
d_{\CP^2}^2 \eea where $dA/2$ gives the K\"ahler class of $\CP^2$.
 As $\psi$ ranges from $0$ to $2\pi$, we get a 3-fold of $\S^5$.
 More generally , we may take an orbifold theory on $\C^3/\Z_m$ where the
generator $g$ of $\Z_m$ acts
 on $\C^3$ by \bea g \cdot (z_1, z_2, z_3) \to (\omega^{a_1},
z_1, \omega^{a_2} z_2, \omega^{a_3} z_3), \omega^m =1, ~~~a_1 + a_2 + a_3 = 0 \pmod{m},a_i >0. \eea Then the
$\S^5$ is a Hopf fibration over a weighted projective space $\CP(a_1,
a_2, a_3)$. As long as
the null geodesic does not lie over the singular locus of the weighted
projective space $\CP(a_1, a_2, a_3)$, there will no change in the
argument.

We now choose the null coordinates as: \bea
\begin{array}{lcl}x^{+}
&= &\frac{1}{2} (t + \frac{1}{3} \psi) \\
x^{-} &= &\frac{R^2}{2} (t - \frac{1}{3} \psi)\end{array} \eea In the
limit $R\to \infty$ and after rescaling
the transversal direction $\CP^2$, we obtain the maximally supersymmetric pp-wave metric
(\ref{ppwaven4}) as in \cite{go}. The light-cone momenta can be written in
terms of conformal weight $\Delta$ and the angular
momentum $J = - i\partial_\psi$: \bea \label{lcmom} \begin{array}{lcl} 2 p^{-} &=& i \partial_{x^{+}} = i
(\partial_{t} + 3
\partial_{\psi}) = \Delta - 3J \\ 2 R^2 p^{+} &= & i\partial_{x^{-}}= i( \partial_{t} -
3\partial_{\psi}) = {\Delta + 3 J}
\end{array}
\eea

Before we describe the duality string/field theory in the Penrose limit, we recall the results of \cite{ks,law}
concerning the field theory on D3 branes at $\C^3/\Z_3$ singularities. By
starting with $3 N$ D3 branes in
the covering space of $\bf{C^3/Z_3}$ orbifold, the $SU(3 N)$ gauge group is broken to $SU(N)^3$ by orbifold
action on the Chan-Paton factors and there are three fields in the bifundamental representation for each pair of
gauge groups, denoted by $X_i, Y_i, Z_i, i=1,2,3$ (they come as 3 $N \times N$ blocks inside each $3 N \times 3
N$ matrices $X, Y, Z$ describing the transversal motion of the D-branes). The surviving KK modes are of the
form~\cite{ot}: \bea \label{az3} \mbox{Tr} (X_i^{m_1} Y_{i+1}^{ m_2} Z_{i+2}^{m_3}),~~~ m_1 + m_2 + m_3 = 0
\pmod{3},~~~ i =1, 2, 3\pmod{3}\eea The quiver gauge theories have a quantum $\Z_3$ symmetry and the surviving KK
modes have to be invariant under it. In the Penrose limit, the effect of the $\Z_3$ action on the transversal
direction to the boosting direction disappears as the string probes  an infinitesimally small neighborhood of the
boosting circle parameterized by $\psi$. In the quantum vacua, the $\Z_3$ action remains along the boosting
direction as we see in (\ref{lcmom}). In the orbifold theory
$\S^5/\Z_3$, the global symmetry $SO(6) \approx SU(4)$ is
broken up  into $U(1) \times \Z_3$. Before the limit, the Hopf fibration
is non-trivial, so even if the $\Z_3$ acts only along the Hopf fiber, this
does not imply the breaking of global
$SO(6)$ isometry. In the pp-limit, the fibration becomes trivial and it
breaks the global symmetry $SO(6)$ to
$SO(4) \times SO(2)$, with the $SO(2)$ being in the boosting direction and $SO(4)$ in the transverse directions.

To describe the string/field theory duality, we denote by $Z$ the
boosted direction and by $X,~Y$ the transverse direction where the
orbifold does not act so  $X,~Y$ do not enter in a gauge invariant
form.~\footnote{ This set of $X,Y, Z$ is different from the original complex coordinates of $\C^3$ in
(\ref{az3}). But a change of complex structures we may identify them as complex coordinates of the infinitesimal
neighborhood of the boosting circle.} The action of $\Z_3$ orbifold is
only on the Hopf fiber parameterized by $Z$. We identify the scalar field
along the Hopf fiber as $Z = Z_1 Z_2 Z_3$ where $Z_i$ are the above fields
in the bifundamental representation of $SU(N)_i \times
SU(N)_{i+1},~i=1,2,3$. The field $Z$ is in the adjoint
representation of $SU(N)$ and has angular momentum on the U(1) direction
equal to $3$. The fields
$X,~Y$ are also in the adjoint representation of the same $SU(N)$ and together with $Z$ they form an
${\cal{N}}=4$ multiplet.

The vacuum of the string in the presence of the $\bf{Z_3}$ is \bea \frac{1}{\sqrt{3 J} N^{3 J/2}}
\mbox{Tr}[Z^{J}] \eea The first excited states are obtained by insertions
of $X, Y, \bar{X}, \bar{Y}$, for string in the
pp-wave background these states being obtained by acting with a single
oscillator on the ground states. Because
there are eight bosonic zero modes oscillators, we expect to find eight bosonic states with $\Delta - 3 J = 1$.
They are \bea \mbox{Tr}[Z^{J}~X],~\mbox{Tr}[Z^{J}~\bar{X}]~\mbox{or}~\mbox{Tr}[Z^{J}~Y],~ \mbox{Tr}[Z^{J}~Y] \eea
and the ones with the covariant derivative \bea \mbox{Tr}[Z^{J}~D_{\mu} Z] \eea The non-supergravity modes are
obtained by acting with creation operators which imply the introduction of a position dependent phase, besides
the above insertions~\cite{bmn}.

Because we discuss the $\bf{Z_3}$ orbifold, we do not have a DLCQ limit as in \cite{mukhi,ali}, which
holds only for $\bf{Z_n}$ with large $n$. Therefore, if we make the identification of the radius of the $x^-$
direction as in \cite{mukhi,ali}: \bea \frac{\pi R^2}{n} = 2 \pi R_{-} \eea where $R^2$ is approximatively $N$
(the rank of the gauge group), we see that when $n$ is small, the radius
$R_{-}$ of the $x^-$ direction is
infinite, so we are not allowed to use a Discrete Light Cone Quantization. There is no winding mode discussion
for the $\bf{Z_3}$ orbifold and the insertions corresponding to the
non-supergravity modes are identical
to the ones of \cite{bmn}.

An interesting case of supersymmetry enhancement was treated in \cite{mukhi,ali} for ${\cal{N}}= 2$ orbifolds
$\bf{S^5/Z_n}$. By boosting along the non-fixed directions of the orbifold, one gets a maximal ${\cal{N}}= 4$
theory. One interesting related development would be to consider the supersymmetry enhancement when D3 branes
probe backgrounds of D7/O7 planes \cite{fay1,fay2}. The Penrose limit in the fixed direction(orthogonal to O7)
was considered in \cite{bmn1} but the discussion of Penrose limits on the non-fixed directions still remains to
be discussed. One step further in this direction would be to consider the Penrose limit for the case when D3
branes probe geometries with orthogonal D7 branes as in \cite{fay2,aot1,asyn}.

\subsection{$\Z_m \times \Z_n$ Orbifolds of $\S^5$}

In this subsection we consider the  geometry $AdS_5 \times \S^5/(\Z_m \times \Z_n)$ which is the near-horizon
limit of the D3 branes placed at the tip of  $\C^3 /(\Z_m \times \Z_n)$ \footnote{This model was also discussed
in \cite{mukhi}}. The coordinates of $\C^3$ are $z1,z_2,z_3$ and the generators $g_m, g_n$ of $\Z_m, \Z_n$ act on
$(z_1, z_2, z_3)$ as \bea
g_m: &(z_1, z_2, z_3) \to (e^{2 \pi i/m} z_1, e^{-2 \pi i/m} z_2, z_3)\\
g_n:  &(z_1, z_2, z_3) \to (e^{2 \pi i/n} z_1, z_2, e^{-2 \pi i/n} z_3). \eea The singular points in the quotient
are points left invariant under elements of the discrete group. The complex curve $z_1 = z_2 = 0$ , parameterized
by $z_3$, is invariant under the $\Z_m$ and becomes a curve of  $\A_{m-1}$ singularities, the complex curve $z_1
= z_3 = 0$ , parameterized by $z_2$, is invariant under the $\Z_n$ and becomes a curve of  $\A_{n -1}$
singularities and the complex curve $z_2 = z_3 = 0$ , parameterized by $z_1$, is invariant under the $\Z_r, r =
\mbox{gcd}(m,n)$ and becomes a curve of  $\A_{r-1}$ singularities.

The field theory on D3 branes at $\C^3/(\Z_m \times \Z_n)$ singularity is ${\cal{N}} = 1$ theory with gauge
group $\prod_{i=1}^{m} \prod_{j=1}^{n} SU(N)_{(i,j)}$ and
chiral bifundamentals \cite{hz,hu}. The gauge invariant
operators are \bea \mbox{Tr} H_{(i,j)(i+1,j)} D_{(i+1,j)(i,j-1)}
V_{(i,j-1)(i,j)} \eea where $H_{(i,j)(i+1,j)}$
are in the bifundamental representation of $ SU(N)_{(i,j)} \times
SU(N)_{(i+1, j)}$, $V_{(i,j)(i,j+1)}$ are in the
bifundamental representation of $ SU(N)_{(i,j)} \times  SU(N)_{(i, j+1)}$
and $D_{(i+1,j+1)(i,j)}$ are in the
bifundamental representation of $ SU(N)_{(i+1, j+1)} \times  SU(N)_{(i, j)}$. If D3 branes move to the points of
 $\A_{m-1}$, (resp. $\A_{n-1}$ or $\A_{r-1}$) singularities described above,
the field theory on the D3 branes becomes ${\cal{N}} = 2$ with gauge
group $SU(N)^{m}$ (resp. $SU(N)^{n}$ or $SU(N)^{r}$). Hence there are flat directions in the ${\CN =1}$ theory
which connect it to an ${\CN =2}$ theory.

In the $\bf{S^5/(Z_m \times Z_n)}$ geometry, there are many interesting directions along which we can consider
the boosting and the amount of the supersymmetry enhancement will depend on both the direction and the locality
of the trajectories. We now classify the different possibilities:

{\bf Case 1.} Boosting in the direction of the $\Z_m$ orbifolding (the same discussion holds for the direction of
the $\Z_n$ or  $\Z_r$ orbifolding).

We understand by the direction of $\Z_m$ orbifolding a $U(1)$ direction
where $\Z_m$ is embedded in.
 For this purpose, it is convenient to consider  $\S^5$ as a foliation of the $\S^3$ in $\C^2$ with coordinates
$z_1, z_2$ and the $\S^1$ in $\C^1$ with coordinates $z_3$. Furthermore, we consider $\S^3$ as a Hopf fibration
over $\CP^1$ after changing the complex structure $z_2$ to $\bar{z_2}$
and the $\Z_3$ will locally act along the
Hopf fiber. From this geometric description of $\S^5$, we obtain the metric for the $AdS_5 \times\S^5$ as:\bea
\begin{array}{cl}
\label{foli1} dS^2_{AdS}  &=R^2(-\cosh^2 \rho dt^2 + d\rho^2 + \sinh^2 \rho d\Omega_3^2)\qquad\qquad\qquad\\
ds^2_{S^5} &=R^2\left[d \theta^2 + \sin^2 \theta d_{\S^1}^2 + \cos^2 \theta \left[\left(d \tau + (\cos \chi
-1)d\phi\right)^2 + \left(d\chi^2 +\sin^2 \chi d \phi^2\right)\right] \right]\end{array}\eea where $\tau$ is the
coordinate for the fiber direction and $d\chi^2 +\sin^2 \chi d \phi^2$ is the metric for the base $\CP^1$ in the
Hopf fibration of $\S^3$. As in the previous section, we need to consider
an $mn$-fold cover of $\S^5$. As the string probes only an infinitesimal
neighborhood of the boosting direction, the action on the transverse
directions to the Hopf fiber is irrelevant. For simplicity we take  an
$m$-fold covering of the $\S^5$ where the $\S^3$ part of the
metric changes to \bea (m~d \tau + (\cos \chi -1)d\phi)^2 + (d\chi^2 +\sin^2 \chi d \phi^2) \eea
 We choose the null coordinates as \bea x^{+} = \frac{1}{2} (t +
\frac{\tau}{m}), \qquad x^{-} = \frac{R^2}{2} (t - \frac{\tau}{m})
\eea and consider a scaling limit $R \to \infty$ around $\theta = \chi = 0$ with \bea \label{n4zm} \rho =
\frac{r}{R}, \,\, \theta = \frac{u}{R},\,\, \chi = \frac{v}{R} \eea In this limit, the metric becomes
\bea\begin{array}{ccl}ds^2 &=& dr^2 + r^2
d\Omega_3^2 -r^2 d{x^{+}}^2- 2dx^{+}dx^{-}\\
&&+du^2 +u^2d_{\S^1}^2 -2 dx^{+}dx^{-} - v^2 dx^{+}d\phi -u^2d{x^{+}}^2 +dv^2 + v^2 d\phi^2\\
&=& -4dx^{+}dx^{-} -(r^2 + u^2){dx^{+}}^2 \\
& &+ dr^2 + r^2 d\Omega_3^2 +du^2 +u^2d_{\S^1}^2+dv^2 + v^2 d\phi^2 -v^2dx^{+}d\phi\\
&=& -4dx^{+}dx^{-} -(r^2 + u^2 +\frac{v^2}{4}){dx^{+}}^2 \\
& &+ dr^2 + r^2 d\Omega_3^2 +du^2 +u^2d_{\S^1}^2+dv^2 + v^2 {d\phi'}^2 \\
\end{array}\eea
where $\phi' = \phi - 1/2x^{+}$.

After changing to the rectangular coordinate system, one may rewrite this as \bea \begin{array}{ccl} ds^2 &=&
-4dx^{+}dx^{-} - \left({\r}^2  + {\u}^2 + {\v}^2 \right) {dx^{+}}^2  +
d{\r}^2 + d{\u}^2 + d{\v}^2
\end{array} \eea
 The pp wave has a natural
decomposition of the $\R^8$ transverse space into $\R^4 \times \bf{R^2} \times \bf{R^2}$ where $\bf{R^4}$ is
parameterized by ${\r}$ and the $\bf{R^2} \times \bf{R^2}$ by ${\u}$ and ${\v}$, respectively. The covariantly
constant flux of the R-R field is on the $(x^+, {\r})$ and $(x^{+}, {\u}, {\v})$. In this geometry, the light
cone momenta are:
\begin{eqnarray}
2 p^{-} = i \partial_{x^{+}} = i (\partial_{t} + m \partial_{\tau}) = \Delta - m J \\ \nonumber 2R^2 p^{+} = i
{\partial_{x^{-}}} = i (\partial_{t} - m \partial_{\tau}) = \Delta + m J \
\end{eqnarray}
The effective angular momentum in the boosting direction is $m J$ and this
is the quantity which should be large in the Penrose limit of the AdS/CFT
correspondence. Therefore we have two options, the first one being to
consider a  discrete orbifold group $\bf{Z_m}$ with a very large $m$ and
finite $J$ and the second a  discrete orbifold group $\bf{Z_m}$ with
finite and with a large value for $J$~\cite{mukhi,ali}.

In the field theory, the supersymmetry is
enlarged from ${\cal{N}} = 1$ to ${\cal{N}} = 4$ and the corresponding
global symmetries are $SO(2)$ in the boosted direction and $SO(4)$ in the
transverse direction to the boosting.
To identify the gauge invariant operators, we need to use the fact
that we boost along the direction of the $\Z_m$ orbifolding and the
rest of the space is invariant.
The direction of the boosting is denoted by $Z$ and, as in the previous
subsection, we denote the transverse coordinates to the boosting by $X$ and
$Y$.  In the terms of the
fields of the  ${\cal{N}} = 1$ theory, $Z$ should be in a gauge invariant
form and is written as a product
$Z = \prod_{i=1}^{m} Z_i$ where $Z_i$ are either  $H_{(i,j)(i+1,j)}$ for
fixed $j$, $D_{(i+1,j+1)(i,j)}$ for fixed $j$ or $V_{(i,j)(i,j+1)}$ for
fixed $i$. The above fields are in the bifundamental
representation of $SU(N)_i \times SU(N)_{i+1}$, the field $Z$ transforms
in the adjoint representation of the group $SU(N)$. Together with the
scalar fields denoting the transverse direction, X and Y, they  form an
${\cal N} = 4$ multiplet.

The coupling of the  $SU(N)$ gauge theory is of order $g_{YM}^2 = g_s m$
and the effective 't Hooft parameter is
$\frac{g_{YM}^2~N}{J^2 m^2}$ which is finite being of the order of
$g_s N m/R^4$ which is finite. Therefore we can treat the $SU(N)$ gauge
theory perturbatively.

We can now proceed to describe the gauge invariant operators corresponding
to the stringy ground state and excitations. The gauge invariant operators
$\mbox{Tr} (Z^J)$ have angular momentum $m J$ in the boosted direction due
to the action of $\Z_m$, and this corresponds to the vacuum of the string
theory. To describe the excitations, we need to consider the two cases
discussed above, i.e. when $m$ is either small or large.

For the case of small $m$ and large $J$, the first level eight bosonic
zero mode oscillators are
\bea
\mbox{Tr} (Z^{J} X),~\mbox{Tr} (Z^{J} Y)~, \mbox{Tr} (Z^{J} \bar{X}),
\mbox{Tr} (Z^{J} \bar{Y})
\eea
together with
$\mbox{Tr} (Z^{J} D_{\mu} Z)$.  In this case $x^-$ is not compact
as is was for the $\S^5/\Z_3$ case discussed in the previous section. The
insertions of $X, Y, \bar{X}, \bar{Y}$ should be made
as $\mbox{Tr} (Z^l X Z^{J-l})$, etc. The non-supergravity
oscillations are obtained by introducing extra phases in the above
operators.

More interesting is the case when $m$ is very large and the light
cone is compact with radius $\frac{\pi R^2}{m}$, the light cone
momentum being quantized $2 p^+ = \frac{m}{R_-}$.  The string theory has a
matrix string description which mimics the one of the flat space as
pointed out in \cite{mukhi,gop}. In \cite{mukhi} string propagation in
DLCQ pp-wave has been considered and the states were labeled by two
quantum numbers, the first being the DLCQ momentum $k$ and the second
being the winding number $m$ in the $x^-$ direction.

The vacuum corresponds to $\mbox{Tr} (Z^J)$ which has $2 p^+ = \frac{m}{R_-}$ and zero winding number. As $J$ is
finite, we can consider $J=1$. The insertions of the fields $X, Y, \bar{X}, \bar{Y}$ should now be made into the
trace of the string of $Z_i$ fields. To do this, we also need to consider the splitting of the matrices $X, Y$
into $m$ $N \times N$ blocks, each one being inserted in $m$ different positions and then a summation over the
position is required to ensure gauge invariance. In terms of the original ${\cal{N}} = 1$ theory, if we chose
$Z_i$ to be the fields $H_{(i,j)(i+1,j)}$ for fixed $j$, then the fields $X$ and $Y$ are build by $m N \times N$
blocks which can be either $D_{(i+1,j+1)(i,j)}$ for fixed $j$ (we denote these by $X_i$) or $V_{(i,j)(i,j+1)}$
for fixed $i$ (we denote these by $Y_i$. By choosing $Z_i$ transform in the bifundamental representation of
$SU(N)_i \times SU(N)_{i+1}$, it results that $X_i$ transform in the bifundamental of $SU(N)_{i+1} \times
SU(N)_{i}$ and $Y^i$ are in the adjoint representation of $SU(N)_{i}$. Therefore the fields $X_i$ should be
inserted in between $Z_i$ and $Z_{i}$ and the fields $Y_i$ should be inserted in between $Z_{i-1}$ and $Z_i$. The
first oscillators with zero winding number will then be \bea \sum_{i=1}^{m}
\mbox{Tr}(Z_1~Z_2~\cdots~Z_{i}~X_i~Z_{i}~\cdots~Z_{m}),  \sum_{i=1}^{m}
\mbox{Tr}(Z_1~Z_2~\cdots~Z_{i-1}~Y_i~Z_{i}~\cdots~Z_{m}), \eea and \bea \sum_{i=1}^{m}
\mbox{Tr}(Z_1~Z_2~\cdots~Z_{i-1}~\bar{X}_i~Z_{i+1}~\cdots~Z_{m}), \sum_{i=1}^{m}
\mbox{Tr}(Z_1~Z_2~\cdots~Z_{i-1}~\bar{Y}_i~Z_{i}~\cdots~Z_{m}), \eea where the summation over $i$ ensures the
gauge invariance. The states which have winding numbers are built with an additional factor $e^{\frac{2 \pi
i}{m}}$ in the above formulas.

In this form, the stringy
operators have an expansion which is similar to the Kaluza Klein expansion
of a generic field of five dimensional theory reduced on a circle used in
\cite{decon,decon1} to conjecture the deconstruction of a five
dimensional theory for large $m$ quiver theories in four dimensions. Our
$\bf{S^5/(Z_m \times Z_n)}$ model should actually be related to a
$(1,1)$ theory in six dimensions \cite{decon1}, but we expect to get a
five dimensions theory as long as we boost along the orbifolding
directions. The two directions needed to deconstruct a six dimensional
theory are obtained in different boosting, one discussed in this
subsection and the other discussed in the next subsection.

The conclusion is that a fast moving
particle in the $\tau$ direction
reduces the gauge group to $SU(N)$ and enhances the supersymmetry from
${\cal{N}} = 1$ to ${\cal{N}} = 4$.

{\bf Case  2.} Boosting in the direction of the fixed locus of the $\Z_m$ orbifolding (the same discussion holds
for the $\Z_n$ or $\Z_r$ orbifolding)

We take the same form of the metric as in (\ref{foli1}), we parameterize
the angle of $\S^1$ by $\psi$, the phase of $z_3$ and we boost along the
$\psi$ direction. Since $\Z_n$ acts on $z_3$, we take an
$n$-fold covering of $\S^5$ replacing $\psi$ by $n\psi$ in the metric. We introduce the null coordinates \bea
x^{+} = \frac{1}{2} (t + \frac{\psi}{n}), \qquad x^{-} = \frac{R^2}{2} (t - \frac{\psi}{n}) \eea and consider a
scaling limit $R \to \infty$ around $\theta = \pi/2$ with \bea \rho = \frac{r}{R}, \,\, \theta -\frac{\pi}{2}=
\frac{u}{R}. \eea The computation is essentially the same as in
\cite{z2}, the transversal
$\S^3$ part of the metric $\left[\left(d \tau + (\cos \chi -1)d\phi\right)^2 + \left(d\chi^2 +\sin^2 \chi d
\phi^2\right)\right] $ is left intact in this limit and hence the $\Z_m$
action remains.

We now denote the scalar field parameterizing the boosted direction by $z_3~=~Z$ and the scalar fields
parameterizing the transverse directions by $z_1~=~X,~z_2~=Y$. The $\Z_m$ discrete group acts now on $X,Y, Z$ as
\bea X~\rightarrow~ e^{2 \pi i/m}X,~Y~\rightarrow~ e^{- 2 \pi i/m}Y ,~Z~\rightarrow~ Z \eea and there is also an
action of $\Z_n$ discrete group on the boosting direction: \bea Z~\rightarrow~e^{-2 \pi i/n}Z \eea Because of the
latter action, the field $Z$ should enter at the power $n$ and this is obtained if we consider that $Z$ is a
product of the ${\cal{N}} = 1$ fields $V_{(ij)(i j+1)}$ for fixed $i$. We introduce the notation \bea
Z^n~=~V_{(i,j)(i,j+1)}~V_{(i,j+1)(i,j+2)}~\cdots~V_{(i,j+n-1)(i,j+n)} \eea where $j = j+n~(\mbox{mod} n)$. The
field $Z^n$ is in the adjoint representation of $SU(N)_{i,j}$ for fixed
$i,~j$. For future use,we  also introduce the notation $Z_j =
V_{(i,j)(i,j+1)}$.

In this case the field theory after the boosting becomes ${\cal{N}} = 2$
$\prod_{i=1}^{m} SU(N)_{i}$, the gauge coupling constants of the gauge
groups are of order $g_{YM}^2 = g_s n$ and the effective 't Hooft
parameters are $\frac{g_{YM}^2~N}{J^2 n^2}$ which are finite being of the
order of $g_s N m/R^4$.

Because of the $\Z_m$ projection, the field $Z$ is actually promoted to a
$m~N~\times~m~N$ matrix, with $m$ $N~\times~N$ blocks, each block being in
the adjoint representation of an $SU(N)_{i,j}$. Together with the
corresponding vectors of  $SU(N)_{i,j}$, they form ${\cal{N}} = 2$
multiplets. The effective angular momentum in the boosting direction
for $\mbox{Tr} Z^J$ being $n J$, we again have two choices, one when $n$
is small and the other when $n$ is big.

Consider first the case when $n$ is small. The vacuum of the string theory
corresponds to the $\bf{Z_m}$ invariant operators:
\bea \frac{1}{\sqrt{m} J} \mbox{Tr}[S^q Z^{nJ}]
\eea
where
$S = (1, e^{2 \pi i/m},..., e^{2 \pi i (m-1)/m})$ denotes the $q-th$ twisted
sector. The oscillations of the string belong to the untwisted modes which
are of
the type
\bea
\label{unt11}
\mbox{Tr} [S^q Z^{nJ} D_{\mu} (Z)]
\eea
and
\bea
\label{unt22}
\mbox{Tr} [S^q Z^{nJ} \chi]
\eea
where $D_{\mu}$ is the covariant derivative and $\chi$ is the
supersymmetric partner of the scalar $Z$.
The scalar fields $X$ and $Y$ are now $m N~\times~m N$ matrices with $m$
$N~\times~N$ extra diagonal blocks denoted by $X_i$ and $Y_i$, each one
transforming in the  bifundamental representation of the group
$SU(N)_{i~j} \times SU(N)_{i+1~j}$
For the twisted sectors we need to consider states built with oscillators
with a fractional moding.  These are obtained by multiplying with
$X$ and $Y$ which are acted upon by the $\Z_m$ group, together with a
position independent phase factor $e^{\frac{2 \pi i}{J} n(q)}$ when
inserting $X, Y$ and $e^{\frac{2 \pi i}{J} n(-q)}$ for
insertions of $\bar{X}, \bar{Y}$.

The discussion changes when $m$ is very large. In this case we have
a compact light cone with radius $\frac{\pi R^2}{n}$ and the light cone
momentum is quantized $2 p^+ = \frac{n}{R_-}$. The
vacuum and the oscillations of the string belonging to the untwisted modes
are the same as before, but we have a change in the definition of the
oscillations of the twisted sectors. The insertions of $X$ and $Y$ should
be now made between $Z_{j-1}$ and $Z_{j}$. To do
this, we have to consider all the blocks $X_i, Y_i, i=1,~\cdots~,m$
as $n N \times n N$ matrices and to split each one of them into $n$
diagonal $N \times N$ blocks denoted by $X_{i~j}, Y_{i~j}, i=1,~\cdots~,n$
for fixed $i$. The insertions will then be
\bea
\sum_{j=1}^{n} \mbox{Tr} (Z_1~\cdots~Z_{j-1} X_{i~j} Z_{j}~\cdots~Z_{n})
\eea
or
\bea
\sum_{j=1}^{n} \mbox{Tr} (Z_1~\cdots~Z_{j-1} Y_{i~j} Z_{j}~\cdots~Z_{n})
\eea
where $j$ denotes the insertion and $i$ denotes the twisted sector.
The winding modes are obtained by using the same formulas with
an extra $e^{\frac{1 \pi i}{n}}$ factor.

The conclusion is that a fast moving particle moving in the $\psi$
direction reduces the gauge group to $\prod_{i=1}^{m} SU(N)_{i}$ and
enhances the supersymmetry from ${\cal{N}} = 1$ to ${\cal{N}} = 2$.

{\bf Case 3.} Boosting in a general direction which is neither  {\bf Case 1} nor {\bf Case 2}.

In this case both discrete groups  $\Z_m$ or  $\Z_n$ are on the
direction of the boosting and the string probes only a small strip along
this direction, therefore there is no orbifold action on the scalar fields
and the result is a maximal supersymmetric Penrose limit.
Because we do not have any orbifold projection
on the three scalar fields $Z, X, Y$, the situation is similar to
moving the D3 brane from the tip of the $\Z_m \times \Z_n$ orbifold in the
bulk, when the supersymmetry is changed from  ${\cal{N}} = 1$ to
${\cal{N}} = 4$. The string/field theory duality then reduces to the one
of subsection 2.1. There is no change in the angular momentum on the
boosted directions due to the orbifolding.

We have identified several boosting directions which imply an enlargement
of supersymmetry. Three directions are along the  $\Z_m$, $\Z_n$
or $\Z_r$ orbifolding which give maximal supersymmetric pp-limits,
three directions are along the fixed loci of $\Z_m$, $\Z_n$ or $\Z_r$
orbifolds which  give ${\cal{N}} = 2$ supersymmetry and an infinite number
of boosting directions are along a general direction which would give
${\cal{N}} = 4$.

The discussion is different for large $m, n$ as compared to the case of
small $m, n$. For the first case we get compact a compact light cone and
this can be used to describe the pp-wave as the limit of a DLCQ theory
with fixed $p^+$. In terms of the choice of boosting, we get a specific
circle so we get a two dimensional torus when both $m$ and $n$ are large.
These two directions are the ones used by \cite{decon1} to describe the
deconstruction of the six dimensional (1,1) theories.

\subsection{Correlation Functions and Supersymmetry Enhancement}
In \cite{minwalla}, a detailed analysis has been made for the anomalous
dimensions and three point functions for the chiral and almost chiral
operators introduced by \cite{bmn} (see also \cite{cor} for similar
discussion). In particular, the authors of \cite{minwalla} have identified
the parameter $g_2 = J^2/N$ as the genus counting parameter in the free
Yang-Mills theory, such the correlation function of
$\mbox{Tr} \bar{Z}^J$ and $\mbox{Tr} ZJ$ has a contribution
$J N^J g_2^{2 h}$ from the genus $h$ Feynman diagram.

We want to see what happens for the  pp-wave limits of orbifold theories. To do this, we start from the
observation that the  correlation functions for the orbifold theories coincide with those of ${\cal{N}} = 4$
theory, modulo the rescaling of the gauge coupling constant, as observed in \cite{ber1} with  string theory
methods and \cite{ber2} by using field theory methods. If we consider the case $\S^5/\Z_n$, in the ${\cal{N}} =
2$ theory we have a factor of $1/n$ in front of the correlations functions. After the Penrose limit on the
non-fixed direction, we go from the orbifolds theory to the covering space and therefore the factor $1/n$
disappears. The correlation functions for the orbifold theories are then expected to have a similar expansion in
genus as in the $\S^5$ case. It would be interesting to show this in
detail, by analogous computations to  \cite{ber1,ber2}.

\section{The ${\cal{N}} = 1$ orbifolds of $\T^{1,1}$}
The case of D3 branes at the conifold or at orbifolds of the conifold
has been discussed extensively in the literature \cite{kw,ura,ot1,gns}.
The
conifold is a three dimensional hypersurface singularity in $\bf{C^4}$ defined by: \bea \label{conifold} z_1 z_2
- z_3 z_4 = 0 \eea which is a metric cone over the 5-dimensional Einstein manifold $\T^{1,1}= SU(2) \times
SU(2)/U(1)$. The conifold can be realized as a holomorphic quotient of $\bf{C^4}$ by the $\bf{C^*}$ action given
by \cite{kw} \bea \label{c-star} (A_1, A_2, B_1, B_2) \rightarrow (\lambda A_1, \lambda A_2, \lambda^{-1} B_1,
\lambda^{-1} B_2). \eea  The map \bea \label{iso} z_1 = A_1 B_1, z_2 = A_2 B_2, z_3 = A_1 B_2, z_4 = A_2 B_1 \eea
provides an isomorphism between these two representations of the conifold. The horizon $\T^{11}$ can be
identified with $|A_1|^2 + |A_2|^2 = |B_1|^2 + |B_2|^2 = 1$ quotient by an $U(1)$ action induced by
(\ref{c-star}). Following \cite{co}, we can parameterize $A_i, B_i$ in terms of Euler angles of $SU(2) \times
SU(2)$: \bea\label{eulerpara}
\begin{array}{ll} A_1 = \cos \frac{\theta_1}{2} \exp {\frac{i}{2} (\psi_1
+ \phi_1)} & A_2 = \sin
\frac{\theta_1}{2} \exp {\frac{i}{2} (\psi_1
-\phi_1)} \\
B_1 = \cos \frac{\theta_2}{2} \exp {\frac{i}{2} (\psi_2 + \phi_2)} & B_2 = \sin \frac{\theta_2}{2} \exp
{\frac{i}{2} (\psi_2 -\phi_2)},
\end{array}\eea and the $U(1)$  is diagonally embedded in $SU(2) \times SU(2)$. After taking a further quotient by
the remaining $U(1)$ factor of $SU(2) \times SU(2)$, we obtain a product
of two projective spaces $\CP^1_1 \times
\CP^1_2$ and may identify the parameters $\theta_i, \phi_i$ with the spherical coordinates of $\CP^1_i$ for $i=1,
2$. Now $\T^{1,1}$ is a $U(1)$ fibration over $\CP^1_1 \times \CP^1_2$ and the $U(1)$ fiber can be  parameterized
by $\psi := \psi_1 + \psi_2$. The Einstein metric on $\T^{1,1}$ of radius
$R$ is
\bea \label{t11metric}
ds^2_{T^{1,1}}&=R^2\left(\frac{1}{9} (d\psi+\mbox{cos}\
\theta_1d\phi_1+\mbox{cos}\ \theta_2d\phi_2)^2 \right. \cr&~~~~~~ \left. +
\frac{1}{6}(d\theta_1^2+\mbox{sin}^2\theta_1 d\phi_1^2+d\theta_2^2+ \mbox{sin}^2\theta_2d\phi_2^2)\right), \eea

Consider an orbifold theory of the conifold where the discrete group $\Z_m
\times \Z_n$ acts on $A_i, B_j$ by
\bea \label{czk} (A_1, A_2, B_1, B_2) \rightarrow (e^{-2 \pi i/m} A_1,
A_2,e^{2 \pi i/m} B_1, B_2), \eea and \bea
\label{czl} (A_1, A_2, B_1, B_2) \rightarrow (e^{-2 \pi i/n} A_1, A_2,
B_1, e^{2 \pi i/n} B_2). \eea

The action (\ref{czk}) descends to the  horizon $\T^{1,1}$ and yields two fixed circles $|A_2|^2 = |B_2|^2 = 1,
A_1 = B_1 = 0$ (mod $U(1)$) and $|A_1|^2 = |B_1|^2 = 1, A_2 = B_2 = 0$
(mod $U(1)$) \cite{ot1}. Similarly, the action
(\ref{czl})  yields two fixed circles $|A_2|^2 = |B_1|^2 = 1, A_1 = B_2 = 0$ (mod $U(1)$) and $|A_1|^2 = |B_2|^2
= 1, A_2 = B_1= 0$ (mod $U(1)$). The horizon $\T^{11}/(\Z_m \times \Z_n)$
is singular along  these
circles, having  an $\A_{m-1}$  singularity along the first two circles
and an $\A_{n-1}$  singularity  along the last two circles. The discrete
group $\Z_m \times \Z_n$ breaks the $SU(2) \times
SU(2)$ part of the isometry group $SU(2)
\times SU(2) \times U(1)$ of $\T^{1,1}$ and  the $U(1)$ part remains as
the global R symmetry.

In terms of Euler angles of $SU(2) \times SU(2)$, the discrete group $\Z_m
\times \Z_n$ action is given by \bea
\begin{array}{lll}(\psi_1, \phi_1, \psi_2, \phi_2)& \to &( \psi_1 -2\pi
i/m, \phi_1 -2\pi i/m, \psi_2 + 2\pi
i/m, \phi_2 +2\pi i/m)\\
(\psi_1, \phi_1, \psi_2, \phi_2)& \to &( \psi_1 -2\pi i/n, \phi_1 -2\pi
i/n, \psi_2 + 2\pi i/n, \phi_2 - 2\pi
i/n)
\end{array}
\eea
What we see from the above equations is that the coordinate of the $U(1)$
fiber ($\psi = \psi_1 + \psi_2$) is left invariant under the action of
$\Z_m
\times \Z_n$, as should be in order to preserve the ${\cal{N}} = 1$
supersymmetry.

Now we study the Penrose limits of $AdS_5 \times \T^{1,1}/\Z_m \times
\Z_n$. The metric for $AdS_5 \times \T^{1,1}$ is
 \bea
\begin{array}{ll}ds^2 &=
R^2[ -\cosh^2 \rho dt^2 + d\rho^2 + \sinh^2 \rho d\Omega_3^2 \\
&\frac{1}{9} (d\psi+\mbox{cos}\ \theta_1d\phi_1+\mbox{cos}\ \theta_2d\phi_2)^2 +
\frac{1}{6}(d\theta_1^2+\mbox{sin}^2\theta_1 d\phi_1^2+d\theta_2^2+ \mbox{sin}^2\theta_2d\phi_2^2)]
\end{array}
\eea
The Penrose limit for the conifold has been studied in
\cite{ikm,go,pando}. As in the previous section, there are many directions
of boosting. We want to study the boosting along the fixed locus of the
discrete group action. Consider first
the boosting along the circle $|A_1|^2 = |B_1|^2 =1, A_2 = B_2 =0$ (mod $U(1)$) which is a fixed locus
of $\Z_m$ action. In terms of the parameters used in (\ref{t11metric}),
this is located at $\theta_1 =\theta_2 =0$
and can be parameterized by $\psi + \phi_1 + \phi_2$. Because of the
action of $\Z_n$, we are actually dealing with an n-covering of $\T^{1,1}$
and the metric of  $\T^{1,1}$ changes into
\bea
\frac{n^2}{9} (d\psi+\mbox{cos}\ \theta_1d\phi_1+\mbox{cos}\
\theta_2d\phi_2)^2 +
\frac{1}{6}(d\theta_1^2+\mbox{sin}^2\theta_1 d\phi_1^2+d\theta_2^2+
\mbox{sin}^2\theta_2d\phi_2^2)]
\eea

We introduce the null coordinates \bea
x^+&=\frac{1}{2}\left(t+{1\over 3~n}(\psi+\phi_1+\phi_2)\right)\cr
x^-&=\frac{R^2}{2}\left(t-{1\over 3~n}(\psi+\phi_1+\phi_2)\right)
\eea and consider a scaling limit $R \to \infty$ around $\theta_1
=\theta_2 = 0$ with \bea \rho = \frac{r}{R},~~~\theta_i =
\frac{\sqrt{6}}{R} \xi_i,~~ i=1,2\eea
 and in the limit $R
\rightarrow \infty$, the metric becomes: \bea
\label{limitc1}
\begin{array}{lcl} ds^2 &=& -4 dx^{+}dx^{-}  - r^2 dx^{+2} +dr^2 +
r^2d\Omega_3^2\\&&+ \sum_{i=1, 2}(d\xi_i^2+ \xi_i^2 d\phi_i^2  - 2 \xi_i^2  d\phi_i d x^+) \\&=& -4 dx^{+}dx^{-}
+ d\r^2  - (\r \cdot \r + \w\cdot\bar{\w }) dx^{+2} + d \w d\bar{\w}
\end{array}
\eea where $\w= (\xi_1e^{i (\phi_1 - x^+)},\xi_2e^{i (\phi_2 -
  x^+)})$.

The same discussion can be extended to the other fixed circles $ A_1 =
B_1 = 0$, $ A_1 = B_2 = 0$ and $ A_2 = B_1 =
0$, where the boosting is again on the direction $\psi+\phi_1+\phi_2$, but
around $\theta_1 = \theta_2 = \pi$,
$\theta_1 = \pi, \theta_2 = 0$ and $\theta_1 = 0, \theta_2 = \pi$,
respectively. The Penrose limit will be
identical with (\ref{limitc1}) after redefining $\w$ appropriately. The transverse  space $\R^8$ decomposes into
a product of $R^4$ which is in the $r^i$ directions and $\C^2$ whose coordinates is given by $w_1$ and $w_2$. We
now investigate the effect of the orbifolding on the geometry of the pp-limit. Note that if we project the
conifold (\ref{conifold}) in $\C^4$ to $\C^3$ by $(z_1, z_2, z_3, z_4) \to (z_1, z_3, z_4)$, we can identify the
the boosting direction of $\T^{1,1}$ with the angular direction of $z_1$ which is parameterized by $1/2(\psi +
\phi_1 + \phi_2)$ as in  (\ref{iso}) and (\ref{eulerpara}), and the
transversal space $\C^2$ can be parameterized by
$z_3$ and $z_4$. On the pp-limit, $\Z_n$ acts on the boosting direction as
\bea z_1 \to e^{-2\pi i/n} z_1 \eea
and on the transversal direction trivially, and on the other hand, $\Z_m$
acts on the transversal direction as
\bea (z_3, z_4) \to (e^{-2\pi i/m} z_3, e^{2\pi i/m} z_4) \eea and acts trivially on the boosting direction $z_1$
which is along the circle of boosting. In terms of the coordinate of the boosting direction,there is an $\Z_n$
action on $\tilde{\psi} = \psi + \phi_1 + \phi_2$ as \bea \tilde{\psi} \to
\tilde{\psi} - \frac{\pi i}{n}. \eea

We now identify the field theories gauge invariant operators which are
dual to the strings modes on the above pp-wave geometry.
The transverse space is
$\bf{S^5}/Z_n$ or
$\bf{S^5}/\bf{Z_m}$, so the field theory is  ${\cal{N}} = 2$,
$\prod_{i=1}^{n} SU(N)_{i}$ or
$\prod_{i=1}^{m} SU(N)_{i}$. Before the boosting the field theory
is ${\cal{N}} = 1$ \bea
\prod_{i=1}^{m} \prod_{j=1}^{n} SU(N)_{ij} \times \prod_{i=1}^{m}
\prod_{j=1}^{n} SU(N)'_{ij}. \eea and there are bifundamental fields
$(A_1)_{i,j;i,j}$ in $SU(N)_{i,j} \times SU(N)'_{i,j}$, $(A_2)_{i+1,j+1;i,j}$ in
$SU(N)_{i+1,j+1} \times SU(N)'_{i,j}$, $(B_1)_{i,j;i,j+1}$ in $SU(N)'_{i,j} \times SU(N)_{i,j+1}$ and
$(B_2)_{i,j;i+1,j}$ in $SU(N)'_{i,j} \times SU(N)'_{i+1,j}$. The products of
$A_i, B_j$, which enter in the definitions of $z_1, z_3, z_4$ are
$(A_1 B_1)_{i,j;i,j+1}$ in  $SU(N)_{i,j} \times SU(N)_{i,j+1}$,
$(A_1 B_2)_{i,j;i+1,j}$ in  $SU(N)_{i,j} \times SU(N)_{i+1,j}$, and
$(A_2 B_1)_{i+1,j+1;i,j+1}$ in  $SU(N)_{i+1,j+1} \times SU(N)_{i,j+1}$.

We want to see the change in the field theory after the boosting. There
are four possible particular cases
of particular gauge groups which correspond to D3 branes at four
dimensional $\A_{m-1}$ or $\A_{n-1}$
singularities, and the field theory becomes  ${\cal{N}} = 2$,
$\prod_{i=1}^{m} SU(N)_{ij}$ or $\prod_{i=1}^{m} SU(N)'_{ij}$  for fixed
$j$ and $\prod_{j=1}^{n} SU(N)_{ij}$ or
$\prod_{j=1}^{n} SU(N)'_{ij}$ for fixed $i$.
The chiral primaries are constructed from sums of gauge invariant products of
chiral superfields, modulo F- and D- flatness condition \cite{gns}. They
are products of the form
$A_{i_1}B_{j_1} A_{i_2}B_{j_2} \cdots A_{i_{mn}}B_{j_{mn}}$, symmetrized
in $A_i$ and $B_j$. For fixed $i$, a particular example of a chiral
primary involving only $A_1$ and $B_1$ fields is:
\bea
\label{chiralp}
\mbox{Tr}((A_1)_{i,j;i,j} (B_1)_{i,j;i,j+1} (A_1)_{i,j+1;i,j+1}
(B_1)_{i,j+1;i,j+2} \cdots (A_1)_{i,j+n-1;i,j+n-1} (B_1)_{i,j+n-1;i j+n})
\eea
where $j + n = j$ (mod n) so the trace is taken over the adjoint
representation of $SU(N)_{i,j}$. The R-charges of the fields $A_i, B_i$
are not changed by the quotienting so the R-charge of the gauge invariant
operator $(\ref{chiralp})$ is $n$.

We now relate the field theory R-charge with the other $U(1)$ charges that
appear in the field theory and geometry.
In the geometry we have two rotation charges for the
$U(1) \times U(1)$ isometry group which are denoted by $J_1$ and $J_2$ and
they are related to the Cartan generators of the $SU(2) \times SU(2)$
global symmetry of the dual superconformal field theory by \cite{ikm,go}:
\bea
J_a=-i{\partial\over \partial \phi_a}_{\big|x^\pm}=-i{\partial\over
\partial \phi_a}_{\big|t,\psi}
+i{\partial \over \partial\psi}_{\big|t,\phi_i}=
Q_a-{1\over 2}R\qquad
a=1,2
\eea
Because of the $\Z_n$ action on the fixed circle the quotiented conifold,
the above relation becomes:
\bea
\label{rel}
n J_a = n Q_a - \frac{R}{2}
\eea
We use the convention that $A_1$ has $Q_1~=~\frac{1}{2}$ and
 $B_1$ has $Q_2~=~\frac{1}{2}$.
In \cite{ikm,go,pando}, the vacuum of the string theory has been
identified
with the state $J_1 = J_2=0$ and the first oscillations of the strings
with $J_1 = \pm 1, J_2=0$ and $J_1 = 0, J_2 = \pm 1$.

Consider now the boosting along $z_1$ direction and we want to identify
the gauge invariant operators which correspond to the string theory ground
state and first oscillation modes. In the case of the conifold, the ground
state $J_1 = J_2=0$ was  identified with the the gauge theory operators:
\cite{ikm,go,pando}:
\bea
\mbox{Tr} (A_1 B_1)^J,
\eea
the first oscillations $J_1 = -1, J_2 = 0$ or $J_1 = 0, J_2 = -1$ were
identified with multiplication by
\bea
A_1 B_2 ~ \mbox{or} ~ A_2 B_1
\eea
and the first oscillations $J_1 = 1, J_2 = 0$ or $J_1 = 0, J_2 = 1$ were
identified  with multiplication by
\bea
A_1 \bar{A}_2 ~ \mbox{or} ~\bar{B}_2 B_1
\eea
where $A_i, B_i$ are all $N \times N$ matrices.
$A_1 \bar{A}_2 ~ \mbox{or} ~\bar{B}_2 B_1$ came from the semi-conserved
currents of the $SU(2)$ groups and were introduced in \cite{cer}. When
there is a quotient action on the $SU(2)$ groups,
$A_1 \bar{A}_2 ~ \mbox{or} ~\bar{B}_2 B_1$ are not invariant so they do not
appear in the spectrum. Because the supersymmetry in Penrose limit
is ${\cal{N}} = 2$, we do not need
the semi-conserved currents to build ${\cal{N}} = 2$ multiplets and we
only need $A_1 B_2 ~ \mbox{and} ~ A_2 B_1$ in order to build the field
theory duals to the twisted sectors of the string theory.

For the quotiented conifold, the matrix $A_1 B_1$ is promoted to a $m~N \times m~N$ matrix which splits into $m,
N \times N$ diagonal matrices in the adjoint representation $SU(N)_{i,j}, i = 1,\cdots,m$, for fixed $j$. The
matrices $A_1 B_2 ~ \mbox{and} ~ A_2 B_1$  become $m~N \times m~N$ matrices which also split into $m$
extra-diagonal $N \times N$ and each block corresponds to  fields transforming in the bifundamental
representation of $SU(N)_{i,j} \times SU(N)_{i+1,j}$. The boosted direction is acted upon by the discrete group
$\Z_n$ so the invariant quantity is a product as in (\ref{chiralp}) with $n$ copies of $A_1$ and $n$ copies of
$B_1$, of the form \bea (A_1)_{i,j ;i,j} (B_1)_{ i,j;i,j+1} \cdots (A_1)_{i,j+n-1;i,j+n-1} (B_1)_{ i,j+n-1;i,
j+n},\eea which is indeed in the adjoint representation of $SU(N)_{i,j}$. Denoting this by $(A_1 B_1)^n$, we see
that the equation (\ref{rel}) implies that it has $J_1 = J_2 = 0$ and it is the ground state of the string. The
vector field for all $SU(N)_{i,j}$ with fixed $j$, together with the field  $[(A_1 B_1)^n]_{i}$ form an
${\cal{N}} = 2$ multiplet. The ground state is given by $m$ mutually orthogonal $\Z_m$ invariant single trace
operators \bea \label{ground} \mbox{Tr} [S^q (A_1 B_1)^{nJ}] \eea where $S$ is defined as $S = (1, e^{2 \pi
i/m},..., e^{2 \pi i (m-1)/m})$ denotes the $q-th$ twisted sector.

The first level untwisted sectors are built with derivatives and descendants of  $(A_1 B_1)^n$ and are of the
form: \bea \label{unt1} \mbox{Tr} [S^q (A_1 B_1)^{nJ} D_{\mu} (A_1 B_1)^n] \eea and \bea \label{unt2} \mbox{Tr}
[S^q (A_1 B_1)^{nJ} \chi] \eea where $D_{\mu}$ is the covariant derivative and $\chi$ is the supersymmetric
partner of the scalar $(A_1 B_1)^n$.

The first level twisted sectors are written with insertions of
$A_1 B_2 ~ \mbox{and} ~ A_2 B_1$, which are
acted upon by $\Z_m$ but are invariant under  $\Z_n$. They have zero
angular momentum in the boosted direction so they are used to build
first level string oscillations. The discussion is similar to the one of
\cite{z2}.

As the effective angular momentum of the string states is $n J$, we again
have the choice of choosing $n$ to be either small or large. For the case
of large $n$, the insertions of $A_1 B_2 ~ \mbox{and} ~ A_2 B_1$  should
be made between different $(A_1)_{i,j ;i,j} (B_1)_{i,j;i,j+1}$. The Penrose limits of quotiented
conifold will then be the limit of a DLCQ theory with constant $p^+$.

 \section{Conclusions}

In this paper we studied the Penrose limits of different
${\cal N} = 1$ orbifold
geometries of $\S^5$ and $\T^{11}$ which lead to supersymmetric PP-wave
backgrounds with
enlarged supersymmetry. We have considered the gauge
invariant chiral operators in the different Penrose limits and we have
identified the string oscillations in terms of the gauge invariant
operators. We discussed the different choices for the rank of the quotient
groups.

\vskip 2cm

{\Large\bf{Acknowledgments}}

We would like to thank Sunil Mukhi for
correspondence, Keshav Dasgupta for comments on the manuscript
and especially Gianguido Dall'Agata for several important discussions.

\end{document}